\documentclass[conference]{IEEEtran}
\IEEEoverridecommandlockouts
\usepackage{cite}
\usepackage{amsmath,amssymb,amsfonts}
\usepackage{algorithm}
\usepackage{algorithmic}
\usepackage{graphicx}
\usepackage{textcomp}
\usepackage{xcolor}
\usepackage{multicol}
\usepackage{subcaption}
\usepackage{mathrsfs}
\usepackage{soul}
\def\BibTeX{{\rm B\kern-.05em{\sc i\kern-.025em b}\kern-.08em T\kern-.1667em\lower.7ex\hbox{E}\kern-.125emX}}

 \usepackage{caption}
  \usepackage{comment}
  \usepackage{lipsum}

\begin{document}

\title{MAP-CSI: Single-site Map-Assisted Localization Using Massive MIMO CSI}

\author{\IEEEauthorblockN{Katarina Vuckovic}
\IEEEauthorblockA{\textit{Department of Electrical}\\ \textit{and Computer Engineering}  \\
\textit{University of Central Florida}\\
Orlando, USA \\
kvuckovic@knights.ucf.edu}
\and
\IEEEauthorblockN{Farzam Hejazi}
\IEEEauthorblockA{\textit{Department of Electrical}\\ \textit{and Computer Engineering} \\
\textit{University of Central Florida}\\
Orlando, USA \\
farzam.hejazi@ucf.edu}
\and
\IEEEauthorblockN{Nazanin Rahnavard}
\IEEEauthorblockA{\textit{Department of Electrical}\\ \textit{and Computer Engineering}  \\
\textit{University of Central Florida}\\
Orlando, USA \\
nazanin.rahnavard@ucf.edu}
\vspace{-20mm}
}

\maketitle

\begin{abstract}
This paper presents a new map-assisted localization approach utilizing Chanel State Information (CSI) in Massive Multiple-Input Multiple-Output (MIMO) systems. Map-assisted localization is an environment-aware approach in which the communication system has information regarding the surrounding environment. By combining radio frequency ray tracing parameters of the multipath components (MPC) with the environment map, it is possible to accomplish localization. Unfortunately, in real-world scenarios, ray tracing parameters are typically not explicitly available.  Thus, additional complexity is added at a base station to obtain this information. On the other hand, CSI is a common communication parameter, usually estimated for any communication channel. In this work, we leverage the already available CSI data to propose a novel map-assisted CSI localization approach, referred to as MAP-CSI. We show that Angle-of-Departure (AoD) and Time-of-Arrival (ToA) can be extracted from CSI and then be used in combination with the environment map to localize the user. We perform simulations on a public MIMO dataset and show that our method works for both line-of-sight (LOS) and non-line-of-sight (NLOS) scenarios. We compare our method to the state-of-the-art (SoA) method that uses the ray tracing data. Using MAP-CSI, we accomplish an average localization error of 1.8 m in LOS and 2.8 m in mixed (combination of LOS and NLOS samples) scenarios. On the other hand, SoA ray tracing has an average error of 1.0 m and 2.2 m, respectively, but requires explicit AoD and ToA information to perform the localization task.
\end{abstract}


\section{Introduction}
\label{Intro}

Although the Global Positioning System (GPS) has been widely used for decades, its availability and accuracy suffer in urban canyons and indoor environments \cite{901174}. Urban environments generally have a dense population and hence many mobile users.
Additionally, Environmental Protection Agency (EPA) reports that we spend $86.9\%$ of our time in indoor areas \cite{klepeis2001national}. Therefore, a large share of users reside in environments where GPS comes short. Furthermore, with the expansion of location-based services and emergence of new technologies such as autonomous vehicles and unmanned aerial systems (UAS) traffic management, the need for highly accurate positioning data is growing \cite{junglas2008location}. GPS cannot always deliver the required level of accuracy as GPS-enabled smartphones are typically only accurate within a $4.9$ m under open sky \cite{GPSacccuracy1}.
Evidently, there is a need for alternative localization technologies. To this end, researchers have been proposing different localization methods. 

\emph{Trilateration} and \emph{triangulation} are two localization techniques that rely on intersections formed by signals coming from multiple anchor nodes (ANs). Several studies have already proposed adapting these techniques for massive Multiple-Input Multiple Output (MIMO) \cite{9097149,4384492, 7577201}. The disadvantage of these techniques is that they cannot be used for single-site localization, where only one AN is used to accomplish localization.  \emph{Fingerprinting} is a technique that consists of geo-tagging communication parameters (e.g.  Receive Signal Strength (RSS)  or Channel State Information (CSI)), followed by a machine learning model that predicts the user's location when presented with a new measurement. Several fingerprinting techniques have been proposed in mmWave Massive MIMO systems that tackle both static and dynamic environments \cite{8307353,vieira2017deep,sun2019fingerprint,9128640, ferrand2020dnn, de2020mamimo, Hejazi2021DyLocDL}. Most recent works use convolutional neural networks (CNNs) to train the fingerprinting model. Using a CNN approach, these techniques reported millimeter range accuracy\cite{Hejazi2021DyLocDL}. The drawback of data-driven approaches is that they require elaborate measurement campaigns to cover every location in the environment. Furthermore, {these methods also require several} hours of training. During these two relatively prolonged procedures, it is probable that the environment changes and the dataset becomes invalid \cite{zafari2019survey}. 

In mmWave systems, the signal propagation is highly directive and only a small number of spatial paths contribute to the received power \cite{8761825}. Some studies have taken advantage of the sparse channel to formulate the localization problem as a compressive sensing  (CS) problem.  These works typically combine channel estimation and localization by using different optimization techniques such as Bayesian Learning \cite{8761825,9154215} or Distributed Compressed Sensing - Simultaneous Orthogonal Matching Pursuit (DCS-SOMP) \cite{8269069,8240645}. In these techniques, CS is first used to extract Angle of Arrival (AoA), Angle of Departure (AoD), and Time of Arrival (ToA) from the sparse received signal. Then, the location is recovered using the estimated parameters. These models require only one transmitter and have been shown to work both for line-of-sight (LOS) and non-line-of-sight (NLOS) scenarios. However, the simulations in these works use very simple models with a limited number scatterers in the environment. To the best of the authors' knowledge, none of the CS techniques have been validated in complex multipath scenarios where there are many scatterers and blockers present in the environment. Examples of {datasets in such environments include}, but are not limited to the DeepMIMO \cite{alkhateeb2019deepmimo} and the ViWi\cite{alrabeiah2020viwi} datasets. 

Map-assisted positioning with angle and time (MAP-AT) \cite{9013365} uses a 3-dimensional (3D) map of the environment and ray tracing to localize the user in a mmWave system using a single base station (BS). Using AoD and ToA, the authors perform a ray tracing exercise to locate the user. They also show that if AoA and ToA are available instead, the rays can be back-propagated to find the user's location. The main issue with this approach is that ray tracing parameters are not explicitly available in practice. The communication system requires additional effort to acquire these parameters which can be a complex and laborious task for the BS. On the other hand, CSI is a common communication parameter often used to quantify the quality of the link and the performance experienced by the user \cite{8395053}. 
Although MAP-AT works well in theory, the proposed map-assisted CSI (MAP-CSI) localization is a simpler and more practical implementation approach. By leveraging the already estimated CSI, we can alleviate the workload at the BS.


In this work, we propose a novel map-assisted localization approach using CSI data. 
CSI, as one of the most fundamental concepts in wireless communication \cite{8395053}, is estimated any time we want to know the channel properties and link quality. In other words, most practical systems perform some sort of channel estimation.  Many different techniques have already been proposed for channel estimation in massive MIMO and any of these techniques can be used to estimate CSI \cite{9165822}.
CSI data preserves all the propagation characteristics of the channel, but 
AoD and ToA cannot be directly extracted from it. Fortunately, using a linear transformation, CSI can be converted to \emph{angle delay profile (ADP)}\cite{8307353}. 
The ADP is interpreted as a visual representation of all distinguishable  paths between the user and the BS \cite{Hejazi2021DyLocDL}. 
In this paper, we demonstrate how AoD and ToA can be recovered from ADP and then be used to superimpose the main propagation rays on a 2-dimensional (2D) environment map to localize the user. In our simulation, we consider a single user and a single BS equipped with a MIMO antenna. We explore two scenarios located in the same environment. In the LOS scenario, all user location data points are in the BS LOS view. While the Mixed scenario has approximately half of the samples with the LOS view obstructed.

The main contributions of our work can be encapsulated as follows: 
\begin{itemize}
    \item Modifying the ADP definition in \cite{sun2019fingerprint} to increase the angular and temporal resolution.
    \item Proposing an algorithm to cluster potential user's locations and select which cluster corresponds to the actual user's location.
    \item Benchmarking the performance of map-assisted localization techniques, MAP-AT and MAP-CSI, on the ViWi public dataset \cite{alrabeiah2020viwi} for both LOS and Mixed dataset.
\end{itemize}

The rest of the paper is organized as follows.
In Section \ref{SystemModel}, we define the channel model and describe how ToA and AoD are obtained from CSI. Next, in Section \ref{Localization}, we discuss both the MAP-CSI and the MAP-AT methods. Then, in Section \ref{Simulation}, we present the {employed} dataset and simulation results. Finally, in Section \ref{Conculsion}, we summarize the work and highlight the main points.  
\section{System Model}
\label{SystemModel}

Consider an environment with a single user and a single BS communicating using a typical MIMO-Orthogonal Frequency Division Multiplexing (OFDM) wireless network. For the ease of exposition, we use the channel model similar to  \cite{ali2017millimeter}. Suppose that the BS is equipped with a uniform linear array (ULA), with half wavelength spacing between two adjacent antennas, and the user's device has a single omni-directional antenna. The BS has $N_t$ antennas, and uses OFDM signaling with $N_c$ sub-carriers. We assume a geometric channel model between the BS and the user with $C$ distinguishable clusters. Moreover, each cluster constitutes of $R_C$ distinguishable paths. Each path can be characterized by a delay, also referred to as ToA ($\tau_{m}^{(k)}$), an AoD from the BS ($\theta_{m}^{(k)}$), and a complex gain ($\alpha_{m}^{(k)}$), where  $k \in \{ 1, \dots,C\}, m \in \{ 1, \dots,R_C\}$ \cite{ali2017millimeter}. Assuming a wide-band OFDM system, we can define the ToA as
\begin{equation}
   \tau_{m}^{(k)} = n_{m}^{(k)} T_s,
    \label{eq_ToA}
\end{equation}
where $T_s$ and $n_{m}^{(k)}$ are the sample interval and the sampled delay belonging to the path $m$ of the cluster $k$, respectively \cite{sun2019fingerprint}. Assuming these parameters, channel frequency response (CFR) for each sub-carrier $l$ can be written as \cite{alkhateeb2016frequency} 
\begin{equation}
    \boldsymbol{h}[l] = \sum_{k=1}^{C} \sum_{m=1}^{R_C} \alpha_{m}^{(k)} \boldsymbol{e}(\theta_{m}^{(k)}) e^{-j 2\pi \frac{l \: n_{m}^{(k)}}{N_c} } \,,
    \label{eq_CSIdef}
\end{equation}

where $j$ is the imaginary unit and $\boldsymbol{e}(\theta)$ is the array response vector of the ULA given by
\begin{equation}
    \boldsymbol{e}(\theta) = [1,e^{-j2 \pi \frac{d cos(\theta)}{\lambda}},\dots,e^{-j2 \pi \frac{(N_t - 1)d cos(\theta)}{\lambda}}]^T\,.
\end{equation}
The overall CFR matrix of the channel, also known as CSI, can be expressed as
\begin{equation}
    \boldsymbol{H} = [\boldsymbol{h}[1],\dots,\boldsymbol{h}[N_c]]\,.
\end{equation}

The ADP is computed from the CSI matrix $\boldsymbol{H}$ as follows
\begin{equation}
   \boldsymbol{A} = \mid \boldsymbol{V}^H \boldsymbol{H} \boldsymbol{F} \mid,
   \label{eq_CSI2ADP}
\end{equation}
where $\boldsymbol{V}$ and $\boldsymbol{F}$ are the discrete Fourier transform (DFT) matrices, $|.|$ denotes absolute value, and $\boldsymbol{V}^H$ is the Hermitian transform of matrix $\boldsymbol{V}$. 
In \cite{sun2019fingerprint}, the DFT matrices are square matrices of size $\boldsymbol{V} \in \mathbb{C}^{N_{t} \times N_{t}}$ and $\boldsymbol{F} \in \mathbb{C}^{N_{c} \times N_{c}}$, which limits the ADP to the size of $\boldsymbol{A} \in \mathbb{C}^{N_{t} \times N_{c}}$. Using this definition, a high-resolution ADP requires a large number of transmitter antenna elements and a large number of sub-carriers. This is often not practical. Here, we change the definition of the ADP to increase the resolution of the ADP without increasing $N_t$ and $N_c$. 
Let us define a new DFT matrix $\boldsymbol{V} \in \mathbb{C}^{N_t \times N_{tt}}$ as
 
 $$[\boldsymbol{V}]_{\: z,q} \overset{\Delta}{=} e^{-j  \pi (z-1) cos( \frac{q\pi}{N_{tt}})},$$ 
 
 and new matrix $\boldsymbol{F} \in \mathbb{C}^{N_c \times N_{cc}}$ as $$[\boldsymbol{F}]_{\: z,q} \overset{\Delta}{=} e^{j 2 \pi \frac{zq}{N_{cc}} },$$

where $N_{tt}$ and $N_{cc}$ are arbitrary integers larger than $N_t$ and $N_c$, respectively. Then, the size of the new ADP matrix is $\boldsymbol{A} \in \mathbb{C}^{N_{tt} \times N_{cc}}$.  By increasing $N_{tt}$  and $N_{cc}$, we can increase the angular and temporal resolutions, respectively. 
 \begin{figure}[H]
 \vspace{-1mm}
    \centering
    \includegraphics[width=3.0in,height=2.5 in]{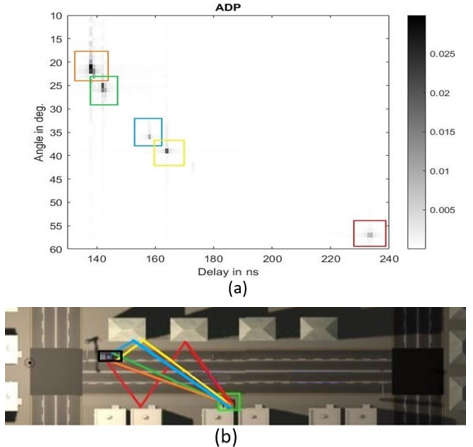}
    \caption {\small {Local maximas in the ADP are marked with the five squares in (a), while their corresponding paths are shown in (b). The green rectangle represents the BS and the black rectangle is the user.  The buildings on both sides of the street are the reflection surfaces. 
    }}
    \vspace{-3mm}
    \label{fig_testADP}
\end{figure}

 An example of the ADP matrix plot is shown in Fig. \ref{fig_testADP}a. The x-axis corresponds to ToA, the y-axis to AoD and the intensity is RSS. Five dominant clusters may be deducted with ToA and AoD approximately at ($137ns$, $23^o$), ($140ns$, $26^o$), ($157ns$, $35^o$), ($162ns$, $40^o$), and ($237ns$, $57^o$).  The clusters formed around the local maximas consist of multipath components (MPCs) that can be categorized by two types of scattering: 1- Specular Components (SpecC), and 2- Dense Multipath Components (DMC). SpecC is a strong specular-like reflection from large physical surfaces. Around a SpecC there are many weaker DMCs with slightly different angles and delays \cite{poutanen2011geometry}.
Fig. \ref{fig_testADP}b shows the 2D bird-view of the environment and the propagation paths of the five clusters marked in Fig. \ref{fig_testADP}a. By knowing the environment map, we can determined where the reflection surfaces are located in the environment. Then, we use this knowledge in combination with the AoD and ToA to plot each propagation path using the ray reflection model defined in Section \ref{ReflectionModel}. The location where the paths intersect is the user's position. The paths can intersect in more than one location; therefore, we use ToA to estimate the length by each ray to eliminate some of the locations where the paths intersect.  
The total distance traveled by the ray using $\tau_{m}^{(k)}$ is calculated as
\begin{equation}
   d = \tau_{m}^{(k)} * c,
   \label{eq_distance}
\end{equation}
where $c$ is the speed of light and $d$ is the distance. Combining the distance traveled with the environment map, we can find the end destination for each ray.  This point becomes a \emph{candidate user's location}.
Some ambiguity is associated with recovering ToA from the ADP matrix. 
Referring to (\ref{eq_ToA}) and (\ref{eq_CSIdef}), if $n_m^{(k)}<N_c$, the delay in ADP is equal to ToA. However, if $n_m^{(k)} \geq N_c$, the delay calculated in ADP is equal to $T_s\times\pmod{(n_m^{(k)}, N_c)}$, which is not the actual ToA. In other words, the actual ToA is the delay obtained from the ADP plus an unknown multiple of $N_c T_s$. Therefore, we have to consider several multiples of $N_c T_s$ for each ray, each of them resulting in a candidate user's location. 
The candidate user's locations from multiple rays form a cluster around the true user's location. The centroid of this cluster becomes the estimated user's location. 



\section{Localization}
\label{Localization}

\subsection{Ray Reflection Model}
\label{ReflectionModel}
\begin{figure}[H]
 \vspace{-1mm}
    \centering
    \includegraphics[width=2.0in,height=1.2 in]{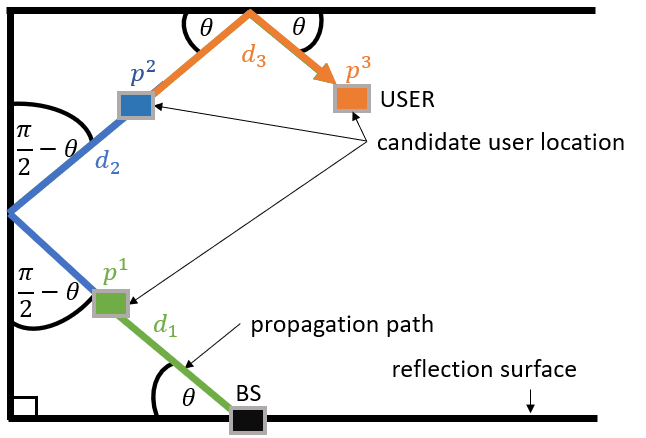}
    \caption {\small {Recursive reflection model. 
    }}
    \vspace{-3mm}
    \label{blocker}
\end{figure}

To model the reflection of the propagation paths, we assume that the reflection surfaces are smooth and use the image-based recursive reflection model defined in \cite{7814249}. The ray is reflected over every surface it encounters until the total distance traveled by the ray is equal to the distance calculated in (\ref{eq_distance}).  An example of the recursive reflection is shown in Fig. \ref{blocker}, where the total distance traveled by the ray is the sum of the $d_i$'s and the AoD is shown as $\theta$. Referring to the ToA ambiguity discussed in Section \ref{SystemModel}, the propagation path shows three candidate user's locations ($p^1$, $p^2$, and $p^3$). The distance $d_1$ is between the BS and $p^1$, where $d_1$ is proportional to distance traveled in time delay calculated from ADP ($n_m^{(k)}<N_c$).  The paths from $p^1$ to $p^2$ and from $p^2$ to $p^3$ are $d_2$ and $d_3$, respectively.  The lengths of $d_2$ and $d_3$ are equal and correspond to the distance traveled in time $N_c  T_s$.


\subsection{Localization using CSI Data}
\label{LocalizationAlg}

Building on the ray reflection model, we propose MAP-CSI localization. 
We start by converting the raw CSI sample, measured at 
time $t$, into an ADP using (\ref{eq_CSI2ADP}). Next, we find the AoD and ToA from the local maximas in the ADP. 
We then combine the (AoD, ToA) pairs with the environment map using the ray reflection model and find the candidate locations.  We label the candidate locations as $p_n^i$ for the $n^{th}$ ray and the $i^{th}$ candidate user's location. Each $p_n^i$ point corresponds to a location in the environment with $x,y$ coordinates.   
Once all $p_n^i$ are collected, the $p_n^i$'s that are outside the area of interest (AoI) are excluded from the set. AoI is the area that encloses all possible user's locations as shown in Fig. \ref{fig_BWlos}.
Typically,  $p_n^i$'s form clusters such that the densest cluster is at the user's location. Based on that, we propose Algorithm \ref{alg2} to identify the clusters and to select which cluster most likely corresponds to the actual user's location.

\begin{algorithm}[H]
\algsetup{
linenosize=\scriptsize,
linenodelimiter=:
}
\vspace{0mm}
\caption{Clustering and Classification}
\begin{algorithmic}[1]

\label{alg2}
\footnotesize
\REQUIRE{collection of possible user's locations $\mathbb{P}$, threshold $D_{th}$, maximum number of clusters $k_{max}$}\\
\ENSURE{estimated number of clusters $k_e$, estimated location $L_e$}
\FOR{all $p_n^i \in \mathbb{P}$}

  \STATE{$s_p \leftarrow$ number of  $p_n^i$ points in $\mathbb{P}$}
        \IF {$s_p$ == 1}
           \STATE{ $k_e = 1$}
           \STATE{$L_e=p_n^i$}
        \ELSE
           \STATE{$\overline{p} \leftarrow \sum_{i=1}^{i_{max}}\sum_{n=1}^{n_{max}} \frac{p_n^i}{i_{max}n_{max}}$} 
           \FOR{all $p_n^i(x,y) \in \mathbb{P}$ }
                \STATE{$D(p) \leftarrow$ distance between $p_n^i$ and $\overline{p} $ }
           \ENDFOR
           \IF{(max($D$)$<D_{th}$)}
                \STATE{ $k_e = 1$}
                \STATE{$L_e = \overline{p} $}
            \ELSE
                \FOR{$k = 2:kmax$}
                    \STATE{calculate $s_n^i(k)$ using (\ref{eq_s}), $\forall i,n$}\\
                    \STATE{calculate $\overline{s}(k)$ using (\ref{eq_smean}), $\forall k$}\\
                    
                \ENDFOR
                \STATE{ calculate $k_e$ using (\ref{eq_sc}) } \\
                \STATE{$ [p_n^i,k_n^i] \leftarrow$ kmeans$(k_{e}$,$\mathbb{P}$), $\forall p_n^i$ }\\
                \STATE{$L_e \leftarrow$ centroid of cluster with most points}\\
            \ENDIF    
      \ENDIF
\ENDFOR \\
\end{algorithmic}
\end{algorithm}
\vspace{0mm}

Algorithm \ref{alg2} has three inputs: maximum number of clusters ($k_{max}$), threshold distance ($D_{th}$) 
 and set $\mathbb{P}$ of points that contains all $p_n^i, \forall n\in[1, n_{max}], i\in[1, i_{max}]$, where $n_{max}$ and $i_{max}$ depend on the environment.
Initially, the algorithm finds the centroid of all points, denoted by $\overline{p}$, and the Euclidian distance of the point farthest from the centroid, denoted by $D$. If $D \leq D_{th}$, then the centroid becomes the estimated location and this means that there exists only one cluster. This is the ideal situation. However, if $D> D_{th}$, this indicates that there are more clusters. In this case, Silhouette Coefficient ($SC$) clustering \cite{kodinariya2013review} is used to find the optimal number of clusters, where $k_{max}$ is the maximum number of clusters considered. The parameters $D_{th}$ and $k_{max}$ are tunable and can vary based on the environment.
The Silhouette value $s_n^i(k)$ is defined as follows 
 \begin{equation}
   s_n^i(k)= \frac{b_n^i-a_n^i}{max(b_n^i,a_n^i)},
    \label{eq_s}
\end{equation}
where $k$ is the number of clusters, $a_n^i$ is the average distance between the point $p_n^i$ and all other points in the cluster to which it belongs to and $b_n^i$ is the minimum of the average distance between point $p_n^i$ and all the points in the other clusters \cite{kodinariya2013review}. The clusters are estimated using k-means. The range of $s$ values is between $-1$ and $1$. If all the $s$'s are close to 1, the set is well classified. On the other hand, if $s$ is close to $-1$, then that point is misclassified. The average Silhouette value for a given $k$ is
 
 \begin{equation}
   \overline{s}(k) =\sum_{i=1}^{i_{max}}\sum_{n=1}^{n_{max}}\frac{s_n^i(k)}{i_{max}n_{max}}.
    \label{eq_smean}
\end{equation}

 After $\overline{s}(k)$ is computed for all $2\leq k\leq k_{max}$, the optimal value ($k_{e}$) is the $k$ corresponding to the maximum $\overline{s}(k)$
 
 \begin{equation}
   k_e =  \arg\max_{k}  \overline{s}(k).
    \label{eq_sc}
\end{equation}

\
Finally, k-means is used to classify all points in $\mathbb{P}$ into $k_{e}$ clusters. Each point 
$p_n^i$ is assigned a class $k_n^i$, where $k_n^i$ ranges from 1 to $k_e$.  The centroid of the cluster with the most points is selected as the estimated user's location $L_e$.  

It is worth noting that both the SC clustering method and k-means Elbow method \cite{kodinariya2013review} were considered when designing the algorithm. Sometimes one method performs better than the other depending on the dataset. However, for this application there was no notable difference between the two methods, so SC was arbitrarily selected. \footnote{The authors release their codes in the following link
”https://github.com/katarinavuckovic/MAPCSI”.}


\section{Simulation Results}
\label{Simulation}
\subsection{Dataset}
\label{Dataset}
\begin{figure}[H]
 \vspace{0mm}
    \centering
    \includegraphics[width= 2.6 in,height= 0.5 in]{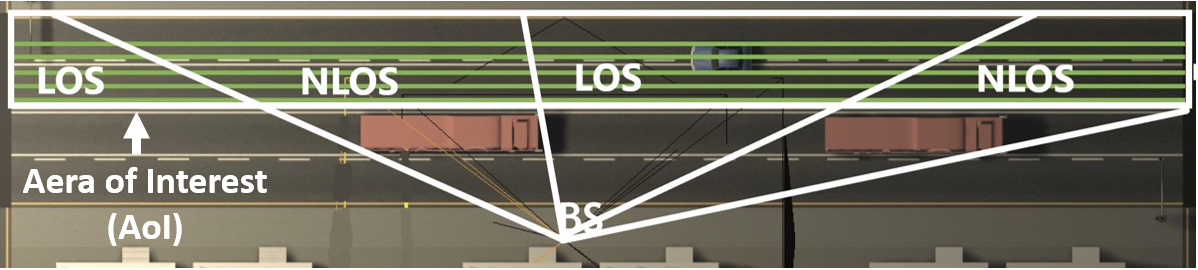}
    \caption {\small {LOS and NLOS regions of the AoI in the Mixed dataset. The red rectangles are the buses that block the LOS view.}}
    \vspace{-3mm}
    \label{fig_BWlos}
\end{figure}

ViWi Dataset \cite{alrabeiah2020viwi} is a public mmWave MIMO dataset. We use datasets from two ViWi scenarios: 1) LOS and 2) Mixed. For the LOS scenario, all locations are in BS LOS view as shown in Fig \ref{fig_testADP}a. On the other hand, the Mixed scenario is created by adding two buses to the same environment.  The buses block the LOS view for approximately half of the samples as shown in Fig. \ref{fig_BWlos}. Furthermore, the AoI is the same for both scenarios.

Every sample in a dataset consists of a user location tagged with AoD, ToA, RSS, and CSI data. The parameters used to generate the dataset are listed in Table \ref{tb1}.  We assume a single BS with an ULA antenna aligned with the $x$-axis with 60 antenna elements. We select the 60 GHz channel. We set OFDM bandwidth to 0.5 GHz and 60 sub-carriers.
\begin{table}[H]
\vspace{-1mm}
 \begin{center}
\caption{\small{Parameters used to generate the Datasets.}}

\begin{tabular}{ |c|c| } 
 \hline
 Frequency Band & 60 GHz\\ 
  \hline
  Bandwidth & 0.5 GHz\\ 
  \hline
Base Station Antenna  & ULA aligned in x-axis \\ 
 \hline
 Antenna Elements ($N_t$) & 60 \\
  \hline
 Sub-carrier Number ($N_c$) & 60 \\
  \hline
   Path Number & 25  \\

\hline
\end{tabular}
\vspace{0mm}
 \label{tb1}
\end{center}  
\end{table}

\begin{figure}[H]
 \vspace{0mm}
    \centering
    \includegraphics[width= 2.8 in,height=0.6 in]{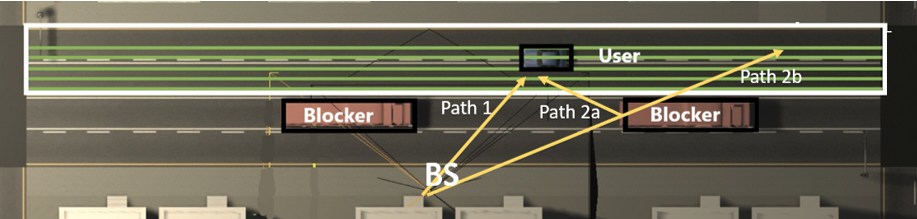}
    \caption {\small {Environment map showing the blockage in Mixed dataset.}}
    \vspace{-3mm}
    \label{fig_BWlabel}
\end{figure}

Fig. \ref{fig_BWlabel} shows the AoI enclosed by the white rectangle that occupies an approximate size of 90 m $\times$ 4 m. There are 5 different horizontal positions (green lines in AoI) and 1000 different vertical positions (not shown), creating a total of 5000 grid points of equally spaced user positions. The buildings and the buses represent reflection surfaces. However, for a path directed towards a bus, there are two viable options that the ray can take. Path 2a is reflected and Path 2b continues traveling in the same direction. We have to consider both of them. The map is the 2D view of the environment and as such does not convey the height of the objects in the image. However, extending this to a 3D view, we realize that the height of the bus is limited and that the ray can still propagate above the bus and reach some of the users located near the NLOS region edges. This creates two different directions that a single ray can propagate in which increases the number of candidate user location points. In our analysis in Section \ref{Results}, we separate the results from the LOS and NLOS regions to analyze the impact of LOS blockage.
The environment also contains some smaller objects such as stop lights and a fire hydrant. We ignore them in our simulations as their reflection surface is hard to model due to their complex shapes. 

\subsection{State of the Art}
We compare our results to the MAP-AT approach, similar to what is presented in \cite{9013365}. In this approach, RSS, ToA, and AoD are required and assumed to be available. To obtain this information, the BS calculates the ToA and AoD for each MPC, which is an oppressive task for the BS. However, assuming the information is available, the authors in \cite{9013365} use multiple MPC components in combination with the environment map to identify candidate locations. 
Here we added the AoI filtering which is not present in the original version of MAP-AT presented in \cite{9013365}. 
AoI filtering is used to improve the performance of the classifier and reduce the number of clusters by discarding the $p_n^i$'s that we know for sure are not at the true user's location. Next, MAP-AT groups  $p_n^i$'s such that the maximum distance between any two points in the cluster is less than $d_{th}$, where $d_{th}$ is a tunable parameter. We modify this part to use Algorithm \ref{alg2} instead to make it easier to compare with MAP-CSI.

\subsection{Results}
\label{Results}
In this section, we compare the results of MAP-CSI to MAP-AT. Although MAP-CSI can never outperform MAP-AT in term of accuracy due to the ambiguity associated with AoD and ToA, it provides a practical implementation solution for a real-world wireless communication system. This is the main advantage of MAP-CSI. Therefore, MAP-AT is presented as the lower bound for the error. We set the tunable parameters in the algorithm to $n_{max}=5$, $i_{max}=7$, and $k_{max}=3$.


\begin{figure}[H]
 \vspace{-1mm}
    \centering
    \includegraphics[width= 3.0 in,height=1.7 in]{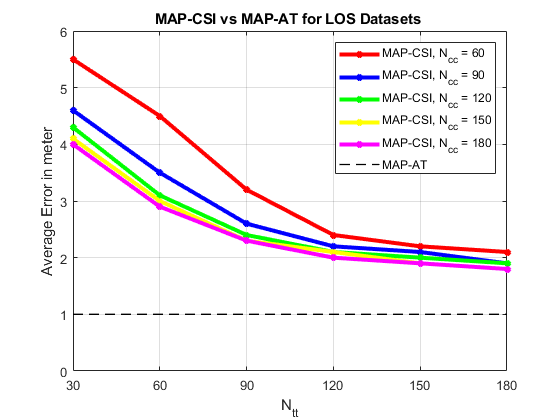}
    \caption {\small {LOS dataset results comparing MAP-CSI and MAP-AT}}
    \vspace{-3mm}
    \label{Fig_results1}
\end{figure}

\begin{figure}[H]
 \vspace{-2mm}
    \centering
    \includegraphics[width=3.0 in,height=1.7 in]{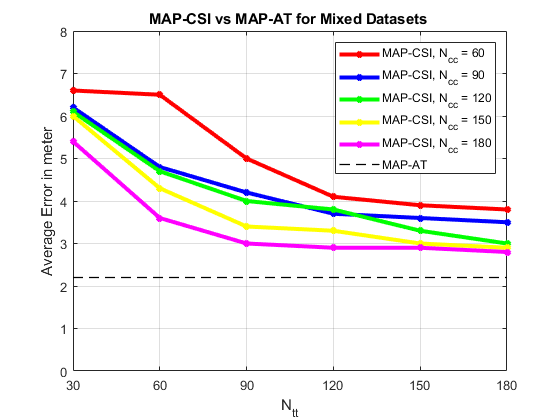}
    \caption {\small {Mixed dataset results comparing MAP-CSI and MAP-AT}}
    \vspace{-3mm}
    \label{Fig_results2}
\end{figure}

Fig. \ref{Fig_results1} and Fig. \ref{Fig_results2} show the average error in meters versus the $N_{tt}$ size for different $N_{cc}$ values for the LOS and Mixed dataset, respectively. The AoD and ToA are directly available in MAP-AT which implies that the size of the ADP is irrelevant and therefore error remains constant for all $N_{tt}$ and $N_{cc}$ values. Furthermore, we observe that increasing $N_{tt}$ and $N_{cc}$ in MAP-CSI reduces the average error. The best results are achieved when  $N_{tt}$ and $N_{cc}$ are both 180 where the error is $1.8$ m for LOS and $2.8$ m for the Mixed dataset. 

\begin{figure}[H]
 \vspace{0 mm}
     \centering
     \includegraphics[width=3.0 in,height=1.5 in]{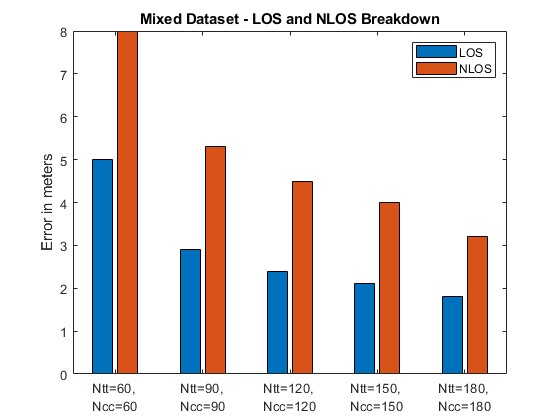}
    \caption {\small {Average Error of LOS and NLOS samples in Mixed Dataset.}}
     \vspace{-3mm}
     \label{Fig_MixedBar}
 \end{figure}

We further examine the Mixed dataset by separating the LOS and NLOS samples as shown in Fig. \ref{Fig_MixedBar}. The error for the LOS samples is comparable to the results in Fig. \ref{Fig_results1}. The error of the NLOS samples is larger since the blockers can obstruct some of the critical paths to the user, thus reducing the number of points in the cluster nearest to the user. Instead, these paths are diverted in other directions forming a denser cluster at another location that is farther away from the user. The classifier may then select the cluster farther from the user as the estimated location causing the error to increase. Nevertheless, the error for both LOS and NLOS samples reduces when $N_{tt}$ and $N_{cc}$ are larger.  The smallest error is achieved when $N_{tt} = N_{cc} = 180$, which is $2.1$ m for LOS region and $3.1$ m for NLOS region.

\section{Conclusion}
\label{Conculsion}
We introduced MAP-CSI, a novel map-assisted localization approach that merely uses the CSI data {and the environment map}. The previous SoA map-assisted technique requires explicit AoD and ToA measurements creating a demanding task for the BS which is nonviable for practical applications. On the contrary, CSI data is usually estimated for any communication channel. We show that AoD and ToA can be estimated from CSI data for every MPC. We compare our results to MAP-AT and show that MAP-CSI can approach MAP-AT accuracy when the size of the ADP is large. 

\vspace{-1mm}
\section*{Acknowledgment}
\vspace{-1.5mm}
This work is supported by the National Science Foundation under Grant No. CCF-1718195.
\vspace{-1.5mm}

\bibliographystyle{IEEEbib}
\bibliography{refs}

\end{document}